\documentclass[longnamesfirst]{emulateapj}

\shorttitle{X-ray Observations of the Bipolar PNe NGC\,2346 and NGC\,7026}
\shortauthors{Gruendl et al.}

\begin{document}

\title{XMM-Newton Observations of the Bipolar Planetary Nebulae
       NGC\,2346 and NGC\,7026}

\author{Robert A. Gruendl\altaffilmark{1}}
\author{Mart\'{\i}n A.\ Guerrero\altaffilmark{2}}
\author{You-Hua Chu\altaffilmark{1}}
\author{Rosa M. Williams\altaffilmark{1}}
\altaffiltext{1}{Astronomy Department, University of Illinois, 
        1002 W. Green Street, Urbana, IL 61801, USA;
        gruendl@astro.uiuc.edu, chu@astro.uiuc.edu, 
        rosanina@astro.uiuc.edu}
\altaffiltext{2}{Instituto de Astrof\'{\i}sica de Andaluc\'{\i}a,
Consejo Superior de Investigaciones Cient\'{\i}ficas (CSIC),
Apartado Correos 3004, E-18080, Granada, Spain; mar@iaa.es}

\begin{abstract}

We have obtained X-ray observations of the bipolar planetary nebulae (PNe) 
NGC\,2346 and NGC\,7026 with {\it XMM-Newton}.  These observations 
detected diffuse X-ray emission from NGC\,7026 but not 
from NGC\,2346.  The X-ray emission from NGC\,7026 appears to be 
confined within the bipolar lobes of the PN and has spectral 
properties suggesting a thermal plasma emitting at a 
temperature of $1.1^{+0.5}_{-0.2}\times10^6$~K.
The X-ray spectrum of NGC\,7026 is modeled using nebular and
stellar abundances to assess whether a significant amount of nebular
material has been mixed into the shocked-wind, but the results of this
comparison are not conclusive owing to the small number of counts detected. 
Observations of bipolar PNe indicate that diffuse X-ray emission
is much less likely detected in open-lobed nebulae than closed-lobed
nebulae, possibly because open-lobed nebulae do not have strong
fast winds or are unable to retain hot gas.

\end{abstract}


\section{Introduction}

Planetary nebulae (PNe) consist of the stellar mass lost by their 
low- or intermediate-mass stellar progenitors during late 
evolutionary stages.
As a star evolves past the asymptotic giant branch (AGB) phase, 
the mass loss process changes from a copious slow AGB wind to a 
tenuous fast wind and the interaction between the fast wind and
the AGB wind forms a PN \citep{K83}.
\citet{FM94} have demonstrated that this wind-wind interaction can
produce the commonly observed elliptical and bipolar morphologies 
of PNe, although the bipolar morphology requires that either the 
AGB wind or the fast stellar wind is anisotropic and has a polar
angle dependence.

The geometry of the fast wind has often been assumed to be isotropic, 
but the presence of bipolar 
jets and collimated outflows in some PNe indicates that the fast 
wind may span a range of geometries.  
For wind velocities greater than $\sim$300~km~s$^{-1}$, the 
interaction of the fast stellar wind with the AGB wind will 
shock-heat the wind material to temperatures greater than 
10$^6$~K.  
This shocked fast stellar wind is too tenuous to produce 
appreciable X-ray emission, but at its interface with the 
dense nebular material, heat conduction lowers the temperature
and mass evaporation raises the density of the shocked fast wind,
producing optimal conditions for X-ray emission \citep{Wetal77,ZP96}.
Since the hot gas in a PN is associated with shocked fast wind,
the distribution of X-ray emission may be used to infer a 
directional dependence for the fast wind.

{\it Chandra} and {\it XMM-Newton} observations of elliptical 
PNe have revealed diffuse X-ray emission confined within their 
innermost nebular shells, e.g., NGC\,2392 \citep{GCGM05} and 
NGC\,6543 \citep{CGGWK01}.  
The plasma temperatures implied by their X-ray spectra are a few 
times $10^6$~K.  In the case of NGC\,6543, the X-ray 
emission is well resolved, and a clear limb-brightening is observed.
Of the ten elliptical PNe observed by either {\it XMM-Newton} or 
{\it Chandra}, six show detectable diffuse X-ray emission
\citep{GCG05}.

{\it Chandra} has observed many bipolar PNe.
While two of them, Mz\,3 and NGC\,7027, exhibit diffuse X-ray 
emission confined within the bipolar lobes \citep{Kast01,Ketal03}, 
the other bipolar PNe (e.g., MyCn\,18, M\,1-16, M\,1-92, M\,2-9, 
OH\,231.8+4.2) were not detected.
Most of these non-detections are understandable because they are
associated with relatively young/exotic bipolar PNe that either 
have large circumstellar absorption or have collimated outflows 
with low wind/outflow velocities, $<$300~km~s$^{-1}$.
In order to better characterize diffuse X-ray emission from 
bipolar PNe, we have used {\it XMM-Newton} to observe two 
carefully chosen bipolar PNe with contrasting lobe morphologies,
NGC\,2346 and NGC\,7026.

NGC\,2346 has a butterfly morphology with two open bipolar lobes 
and a bright equatorial waist that contains large amounts of 
molecular gas and dust \citep{WMW91,ZG88,Setal04}.  
Its central star is a white dwarf with an A5 companion
\citep{M78}, forming a single-line spectroscopic binary 
with a period of $\sim$16 days \citep{MN81}.
Its nebular gas shows a bipolar velocity field with 
a dynamic age of 3,500--4,700 yrs \citep{WMW91,Aetal01}.
NGC\,2346 can be described as an evolved bipolar PN.  

NGC\,7026 has closed bipolar lobes emanating from a bright 
ring-like waist \citep{SW84}, and the lobes appear to have
multiple components \citep{CPM96,L03}.  
Its central star is a hydrogen-deficient WC star
\citep{Aller76,AN03}.
The dynamic age derived from the nebular expansion, $<$1,000 yrs,
suggests that NGC\,7026 is a young bipolar PN \citep{SW84}.

The {\it XMM-Newton} observations of these two PNe have detected 
diffuse X-ray emission from NGC\,7026, but not from NGC\,2346.  
In this paper, we describe these X-ray observations and their 
reduction in \S2, report our analysis of the observations in \S3,
and discuss their implications in \S4.

\section{Observations}

NGC\,2346 and NGC\,7026 were observed with {\it XMM-Newton}
in Revolution 876 (Obs ID 0200240501) on 2004 September 20 and 
Revolution 825 (Obs ID 0200240101) on 2004 June 10, respectively.
The European Photon Imaging Camera (EPIC), which consists of 
two MOS and one PN CCD arrays, was used with a medium filter.
The two EPIC/MOS cameras were operated in the Full-Frame Mode 
for a total exposure time of 16.5 ks for NGC\,2346 and 
19.2 ks for NGC\,7026.
The EPIC/pn camera was operated in the Extended Full-Frame Mode 
for a total exposure time of 11.1 ks for NGC\,2346 and 14.1 ks for 
NGC\,7026.  

The {\it XMM-Newton} pipeline products have been processed using the 
{\it XMM-Newton} Science Analysis Software (SAS version 6.1.0) and the 
calibration files from the Calibration Access Layer available on 2005 
April 22.  
The event files have been screened to eliminate bad events and periods of 
high background.  
For the EPIC/MOS observations, only events with CCD patterns 0--12 
(similar to $ASCA$ grades 0--4) were selected;  
for the EPIC/pn observation, only events with CCD pattern 0 (single 
pixel events) were selected.  
To assess the background rate, we obtained light curves for each 
instrument in the 10--12 keV energy range, where the counts are 
dominated by the background. 
The time intervals with high background, i.e., count rates 
$\ge 0.25$ cnts~s$^{-1}$  for the EPIC/MOS or $\ge 1.0$ 
cnts~s$^{-1}$ for the EPIC/pn, were discarded.
The resulting exposure times for the EPIC/MOS and EPIC/pn observations
were 16.5 ks and 10.9 ks for NGC\,2346 and 15.5 ks and 13.3 ks for
NGC\,7026, respectively.

\section{Results}

X-ray emission was clearly detected from NGC\,7026 but not from
NGC\,2346.  We therefore describe the results from NGC\,7026 first 
and then determine an upper limit for the X-ray luminosity of NGC\,2346.

\subsection{NGC\,7026: Spatial Properties}

To measure the faint X-ray emission from NGC\,7026, we used a
$60\arcsec \times 40\arcsec$ elliptical source aperture that 
encompasses the entire optical nebula and a background aperture 
$\sim$5 times as large as the source aperture located at a similar 
Y-coordinate of the detector.
The EPIC/pn observation detected a total of $110\pm15$
background-subtracted counts, corresponding to a 
count rate of 8.3$\times$10$^{-3}$ cnts~s$^{-1}$.
The EPIC/MOS1 and EPIC/MOS2 observations detected $26\pm8$ and
$31\pm8$ background-subtracted counts, 
corresponding to count rates of 1.7$\times$10$^{-3}$ cnts~s$^{-1}$ 
and 2.0$\times$10$^{-3}$ cnts~s$^{-1}$, respectively. 

We have merged the EPIC/pn and EPIC/MOS images to increase the 
signal-to-noise ratio.  
The merged EPIC image is displayed in Figure~1a and a smoothed 
image in Figure~1b. 
Diffuse X-ray emission is clearly detected.
In order to compare the distribution of X-ray emission with the
optical nebula, accurate alignment between X-ray and optical
images is essential.
We have determined the astrometric solution for the EPIC image 
with an accuracy of $\sim$0\farcs5, based on stellar sources 
that are detected in both this X-ray image and the {\it Digitized 
Sky Survey}.  The relative alignment of the EPIC image and a Nordic 
Optical Telescope (NOT) [\ion{N}{2}] image in Figure~1c is 
accurate to $\sim$0\farcs6.

\begin{figure*}[!t]
\centering
\epsscale{0.9}
\plotone{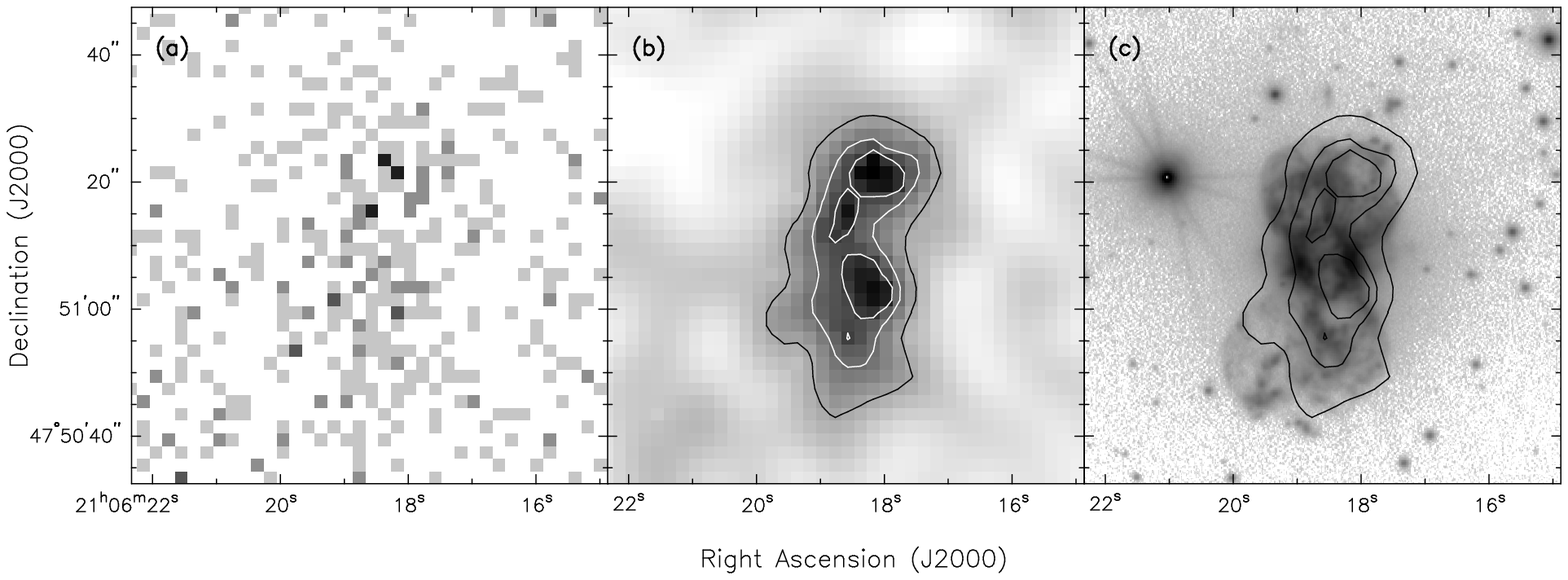}
\vspace{0.25in}
\epsscale{0.8}
\plotone{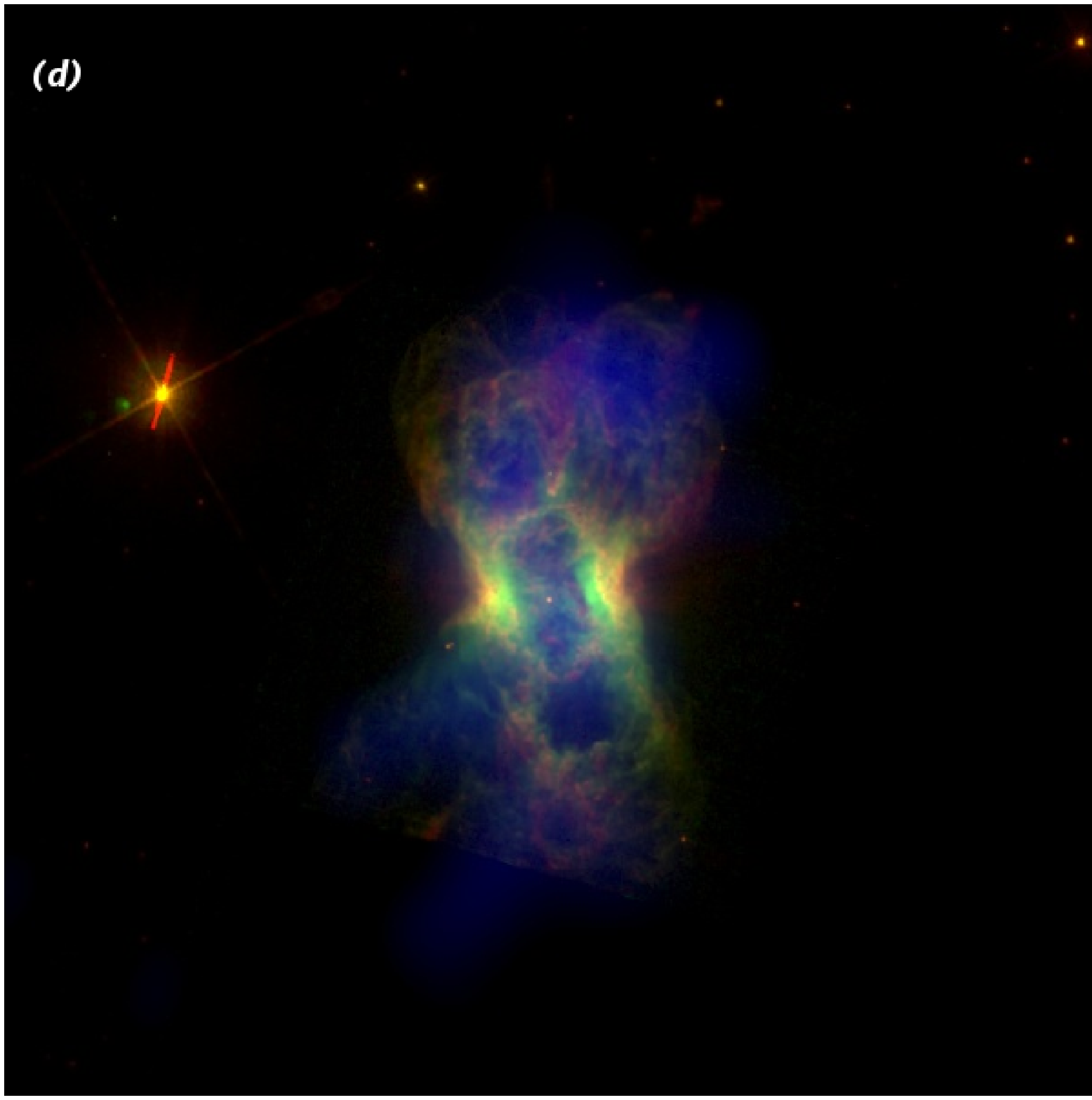}
\vspace{0.25in}
\notetoeditor{The file f1d.eps is color postscript and is intended for 
the electronic edition only.  It is included as a separate panel below.}
\caption{
{\it XMM-Newton} and NOT [N~{\sc ii}] images of NGC\,7026.  Panel {\it (a)} 
displays the {\it XMM-Newton} EPIC raw image in the 0.4-1.1 keV band with 
pixel size 2\farcs0.  This energy band includes most of the photons detected 
from NGC\,7026 as illustrated by its EPIC/pn X-ray spectrum shown in Fig.~2.   
Panel {\it (b)} shows a grey-scale presentation of an adaptively smoothed 
EPIC image using Gaussian profiles with FWHM ranging from 1\farcs5 to 4$''$.  
The X-ray contours overplotted correspond to 50\%, 75\%, and 95\% of the peak 
level.  Panel {\it (c)} displays the NOT [N~{\sc ii}] image overplotted by 
these same X-ray contours.   Panel {\it (d)} shows a color-composite image 
comprised of the smoothed {\it XMM-Newton} EPIC image (blue) and {\it HST} 
images obtained with the F502N (green) and F658N (red) filters to show 
[\ion{O}{3}] and [\ion{N}{2}] emission.
}
\label{img7026}
\end{figure*}
 
The distribution of diffuse X-ray emission from NGC\,7026 
is elongated in the North-South direction, matching the 
polar axis of the nebula.
The position and orientation of the 50\% contour of the smoothed EPIC 
image follow the bipolar lobes shown in the NOT [N~{\sc ii}] 
image (Fig.~1c).  
Furthermore, the X-ray emission shows two peaks that are projected 
within the bipolar lobes north and south of the central waist
of NGC\,7026.  
These excellent correspondences suggest that the diffuse X-ray 
emission from NGC\,7026 is all confined within its bipolar lobes, 
similar to what is seen in the bipolar PN Mz\,3 \citep{Ketal03}.  

We have simulated the surface brightness of diffuse X-ray emission 
assuming the hot gas is distributed in two spherical shells with 
fractional shell thickness ($\Delta R/R$) of 0.1, 0.2, 0.5, and 1.0.
We find that the small number of counts and limited angular resolution
precludes even a rough estimate of the geometry.
Similarly, we cannot determine whether the local minimum along the 
waist is caused by higher circumstellar absorption.

\subsection{NGC\,7026: Spectral Properties}

\begin{figure*}[!th]
\epsscale{1.}
\plottwo{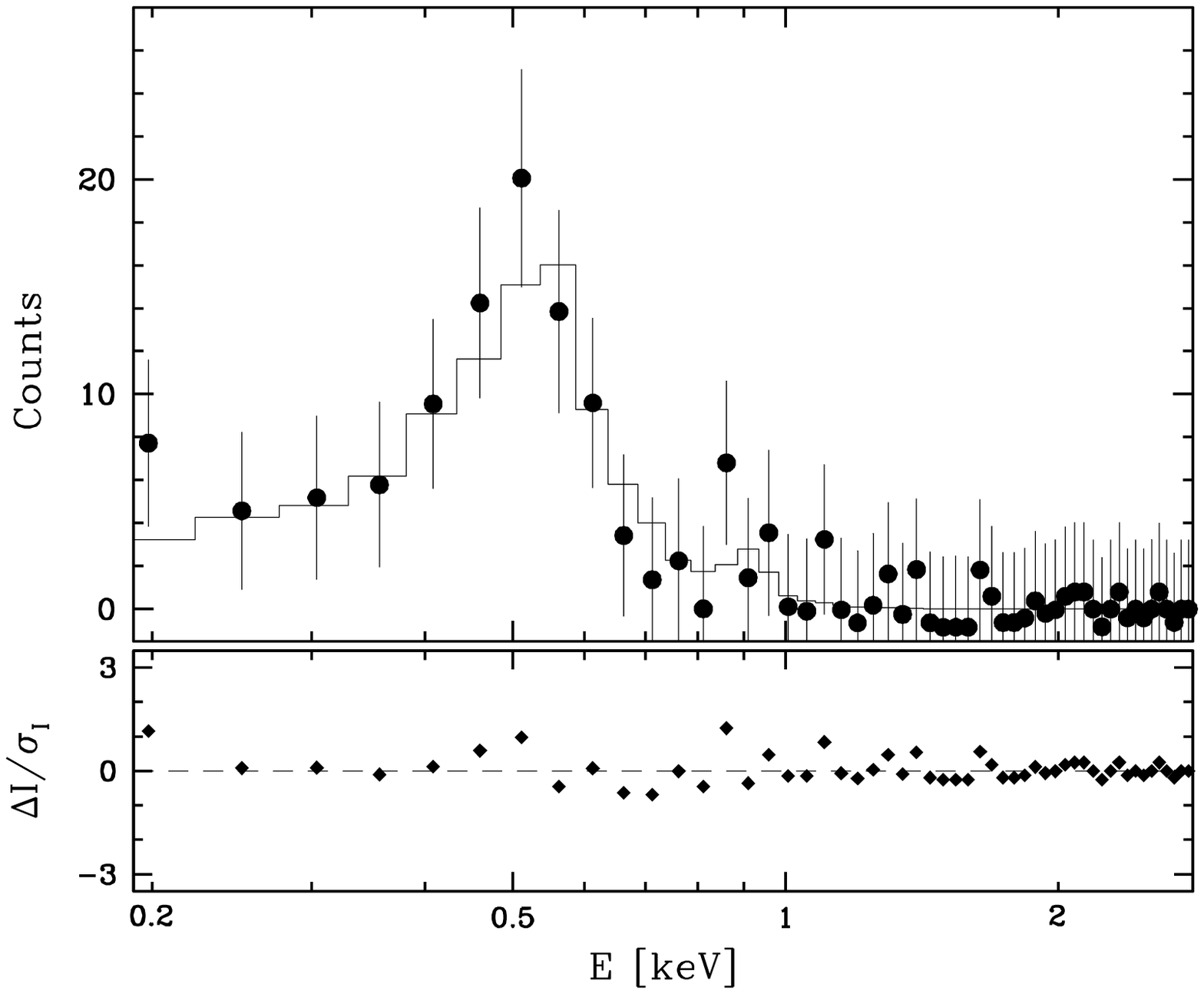}{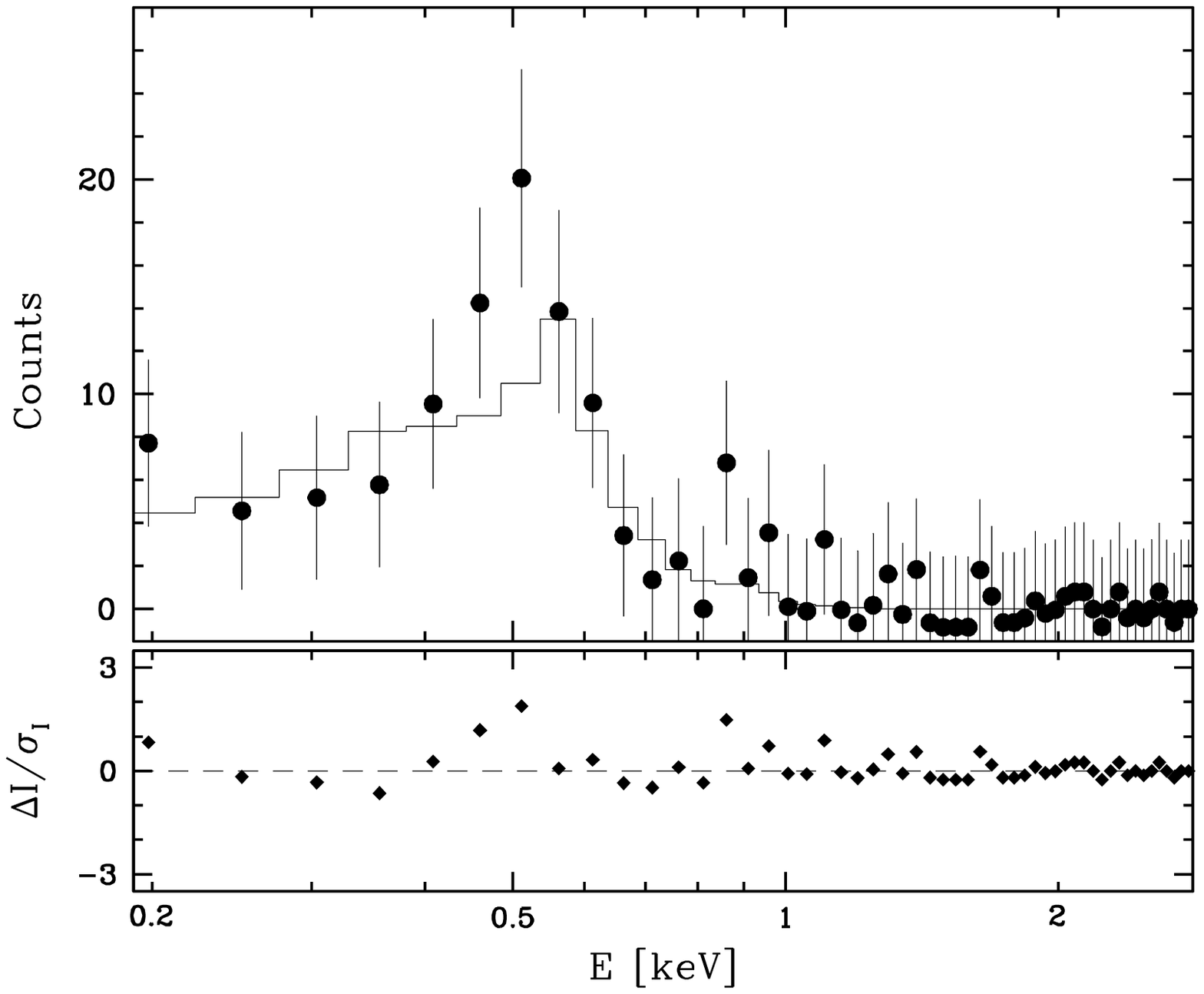}
\caption{
({\it left})
{\it XMM-Newton} EPIC/pn background-subtracted spectrum of NGC\,7026 
overplotted with the best-fit MEKAL model assuming nebular abundances
(solid line).  ({\it right}) the same spectrum overplotted with the best-fit
MEKAL model assuming the abundances of the central star.  
For plotting purposes, both the spectrum and model are binned in 50 eV 
wide energy bins to preserve the EPIC/pn spectral resolution of 
$\sim100$ eV at 1 keV.  Beneath each spectrum the residuals from the best-fit
are plotted.
}
\label{spec7026}
\end{figure*}

X-ray spectra of NGC\,7026 have been extracted from the event files of the 
three {\sl XMM-Newton} EPIC cameras using the same source and background
apertures described in \S3.1.
Due to the limited number of counts detected by the EPIC/MOS 
observations, we will limit the spectral analysis 
to the EPIC/pn spectrum shown in Figure~2.  
This spectrum is soft, peaking at $\sim0.5$ keV with a nearly
flat distribution from 0.2 to 0.4 keV.  Emission between 0.7 and 
1.1 keV is detected but with low S/N ratio.
There is no detectable emission at energies greater than 1.25 keV.  

The EPIC/pn spectrum of NGC\,7026, similar to other PNe, is 
suggestive of thermal plasma emission with its broad peak at 
$\sim$0.5 keV corresponding to emission from the He-like triplets of 
N~{\sc vi} at $\sim$0.43 keV and O~{\sc vii} at $\sim0.57$ keV.  
In order to model this X-ray spectrum, we have adopted the MEKAL thin 
plasma emission model \citep{KM93,LOG95}.
Spectral fits were carried out by folding the thin plasma emission 
model spectrum through the EPIC/pn response matrix, and comparing the 
modeled spectrum to the observed EPIC/pn spectrum in the 0.25--1.5 keV 
energy range after the spectrum was resampled to energy bins with at 
least 16 counts.  
The best-fit was judged based on the $\chi^2$ statistics. 
It was found that the spectral fits did not accurately constrain
the absorption column density, $N_{\rm H}$, if it was allowed to vary
as a free parameter.  We therefore held $N_{\rm H}$ fixed at 
3.2$\times$10$^{21}$~cm$^{-2}$, estimated from the observed logarithmic 
extinction in the H$\beta$ line $c({\rm H}\beta)=0.8\pm0.1$ 
\citep{KH01,HF04} and the canonical gas-to-dust ratio of 
$N_{\rm H}/E(B-V)=5.8\times10^{21}$~atoms~cm$^{-2}$~mag$^{-1}$
\citep{BSD78}.
The X-ray-emitting gas consists of shocked fast wind and mixed-in nebular
gas.  As the wind abundances determined from UV spectroscopic observations
\citep{K01,Hetal05} differ from the nebular abundances determined from 
optical spectra \citep{KH01}, we have made spectral fits using wind or 
nebular abundances separately.

The first spectral fits use the nebular abundances, with the number ratio
of He, O, N, Ne, Ar, Cl, and S to hydrogen being 0.14, 7.3$\times$10$^{-4}$,
5.5$\times$10$^{-4}$, 1.8$\times$10$^{-4}$, 3.1$\times$10$^{-6}$,
3.7$\times$10$^{-7}$, and 1.4$\times$10$^{-5}$, respectively \citep{KH01}.
The corresponding nebular abundances of these elements relative to the 
solar values \citep{AG89} are 1.43, 0.86, 4.9, 1.8, 1.5, 2.0, and 0.96,
respectively.  Solar abundances are used for the other elements, which
have low abundances and are not expected to contribute significantly to
the diffuse X-ray emission.   We have also assumed solar abundances for
the foreground interstellar absorption, and adopted absorption 
cross-sections from \citet{BM92}.
The resulting best-fit model with nebular abundances and a fixed $N_{\rm H}$ 
is overplotted on the EPIC/pn spectrum in Figure~2a.  The plasma temperature 
of the best-fit model is $T = (1.1^{+0.5}_{-0.2})\times10^6$ K 
(i.e., $kT$ = 0.095$^{+0.045}_{-0.015}$ keV).

The second spectral fits use stellar abundances.  The central star of 
NGC\,7026 has been classified as a hydrogen-deficient WC star with the
mass-fractions of He, C, and O being 70, 20, and 10\%, respectively 
\citep{K01,Hetal05}.  In order to make spectral fits using a MEKAL 
model we have assumed a He/H ratio of 200 and scaled the abundances of 
C and O according to their mass fractions.  Thus, the values adopted for
the stellar wind abundances relative to the solar values were 20, 500, 
and 80 for He, C, and O, respectively.  All other abundances were set to 
zero and the absorption column density, $N_{\rm H}$, was again held fixed 
with a value of 3.2$\times$10$^{21}$~cm$^{-2}$.  
The resulting best-fit model is shown in Figure~2b.  This model with stellar
abundances has roughly the same temperature as the best-fit using nebular 
abundances and a slightly higher reduced $\chi^2$.
The most obvious difference between the best-fit models using nebular and 
stellar abundances is that the model with stellar abundances 
underestimates the emission at $\sim$0.5~keV due to the absence of N 
in the stellar chemical composition.
Clearly the quality of both spectral fits are limited by the relatively small 
number of counts in the X-ray spectrum and it is not possible to 
convincingly distinguish whether the chemical composition of the X-ray 
emitting gas in NGC\,7026 better matches that of the central star or 
the nebula.

From the EPIC/pn spectrum we can also determine that the volume emission 
measure is 
$3.6^{+1.4}_{-0.8}\times10^{55} (\frac{d}{1700~{\rm pc}})^2$ cm$^{-3}$, 
where $d$ is the distance to NGC\,7026 in pc.  The distance has been 
estimated to be 
1700$^{+400}_{-200}$ pc based on the extinction to NGC\,7026 \citep{HF04} 
and on the extinction-distance relationship derived by \citet{SW84}.  
The observed (absorbed) X-ray flux of NGC\,7026 in the 0.2--2.5 keV 
energy range is $(8.8\pm1.2)\times10^{-15}$ ergs~cm$^{-2}$~s$^{-1}$, 
and the intrinsic (unabsorbed) X-ray flux is 
$(1.3^{+0.5}_{-0.4})\times10^{-12}$ ergs~cm$^{-2}$~s$^{-1}$. 
The total X-ray luminosity in the 0.2--2.5 keV energy range is 
$L_{\rm X} = (4.5^{+2.0}_{-1.6})\times10^{32} (\frac{d}{1700 {\rm pc}})^2$ 
ergs~s$^{-1}$.

\subsection{NGC\,2346}

The {\it XMM-Newton} observations of NGC\,2346 are compared
with a {\it Hubble Space Telescope} ({\it HST}) WFPC2 [N~{\sc ii}] image 
in Figure~3.  
There is no statistically significant X-ray emission associated with 
either the nebula or the central star in these observations.
The EPIC/pn observations have been used to derive an upper limit for its 
X-ray luminosity using a source aperture encompassing the optical
nebula.  The 3$\sigma$ upper limit for the EPIC/pn count rate in 
the 0.3-1.25 keV band is 4.5$\times$10$^{-3}$ cnts~s$^{-1}$.  
To derive an upper limit of the intrinsic (unabsorbed) flux we assume a thin 
plasma emission model with similar temperature and nebular abundances 
to the one that describes the X-ray emission from NGC\,7026 (see \S3.2).
The foreground absorption is estimated based on the observed logarithmic
extinction at the H$\beta$ line, c(H$\beta$)=0.75 \citep{Cetal99}, which
corresponds to a column density of $N_{\rm H}$=3.0$\times$10$^{21}$ cm$^{-2}$.
At a distance to NGC\,2346 of 690 pc \citep{T97}, the 3-$\sigma$ 
upper limit of the unabsorbed flux corresponds to an X-ray luminosity 
$<$4$\times$10$^{31}$ ergs~s$^{-1}$.

\begin{figure*}[!t]
\centering
\epsscale{0.9}
\plotone{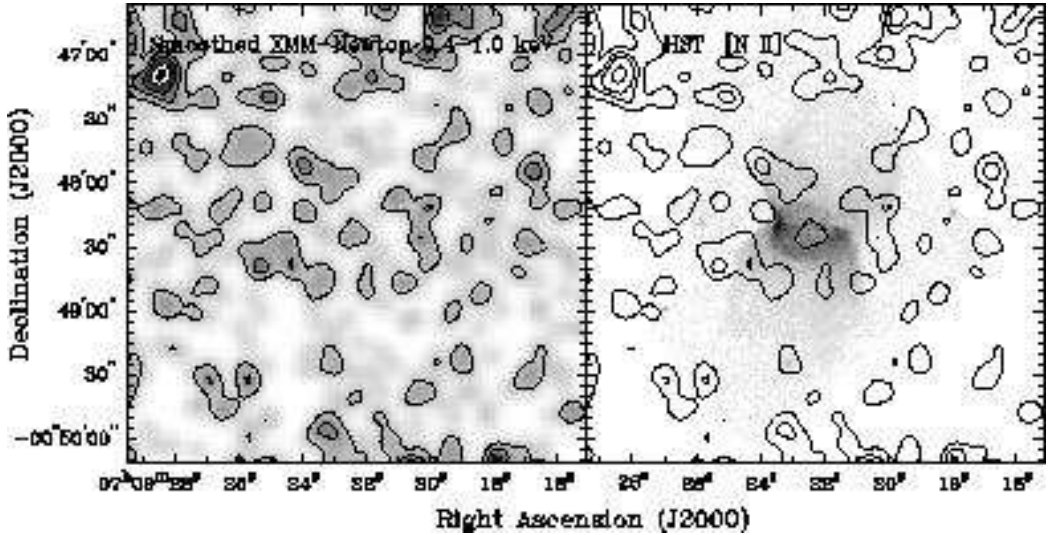}
\caption{
{\it XMM-Newton} and {\it HST} [N~{\sc ii}] images of NGC\,2346.  ({\it left})
{\it XMM-Newton} EPIC image in the $0.25-1.5$ keV band with pixel size 2\farcs0.  
The EPIC/pn and EPIC/MOS images have been merged and the resulting image smoothed using 
Gaussian profiles with FWHM 4$''$ to increase the signal-to-noise
ratio.  The X-ray contours overplotted corresponds to the 1, 3, 5, and 9-$\sigma$ 
over the background.  ({\it right}) {\it HST} WFPC2 [N~{\sc ii}] image 
overplotted by the same X-ray contours.  
}
\label{img2346}
\end{figure*}

\section{Discussion} 

The {\it XMM-Newton} observations of NGC\,7026 clearly show diffuse 
X-ray emission while the observations of NGC\,2346 do not detect significant 
X-ray emission.  We will now use these observations to discuss the 
production and evolution of hot gas in bipolar PNe.

\subsection{NGC\,7026 versus NGC\,2346}

The spatial distribution of the X-ray emission in NGC\,7026 suggests 
that the X-ray-emitting gas is confined within the bipolar lobes, but
its spatial distribution is unknown.  Therefore, we
define the hot gas filling factor, $\epsilon$, as the ratio of the 
emitting volume to the total nebular volume and use a reference value
of 0.5 in our analysis.
Using the volume emission measure determined from the spectral fits
(in \S3.2) and approximating the emitting volume by a 
40$''$$\times$15$''$$\times$15$''$ rectangular box, the rms electron 
density is $\sim22\,(d/1700 {\rm pc})^{-1/2} (\epsilon/0.5)^{-1/2}$ 
cm$^{-3}$.
This density and the best-fit plasma temperature can in turn be used
to derive the thermal pressure of the hot gas, 
$P_{\rm th} \sim 1.9\times n_{\rm e}kT 
\sim6\times10^{-9}~(d/1700 {\rm pc})^{-1/2}~(\epsilon/0.5)^{-1/2}$ 
dynes~cm$^{-2}$, assuming H and He are fully ionized and that the 
number ratio of He/H is 0.14 (the nebular abundance).  

We can compare this hot gas pressure to the thermal pressure of
the warm nebular shell of NGC\,7026 to see whether the nebular 
expansion is driven by the thermal pressure as expected by the
\citet{Wetal77} model.
The electron density and temperature reported by \citet{KH01}
are generally lower than those reported by \citet{HF04}.
The lowest temperature and density are derived from the [\ion{S}{2}]
lines, 6,400 K and 3,300 cm$^{-3}$ \citep{KH01},
resulting in a thermal pressure of $6.3\times10^{-9}$ dynes~cm$^{-2}$.
The temperatures and densities derived from a variety of lines
by \citet{HF04} are $\sim9,000$ K and $\sim10^4$ cm$^{-3}$,
resulting in a thermal pressure of $\sim2.3\times10^{-8}$ 
dynes~cm$^{-2}$.
These measurements were made at the bright nebular waist, where 
the density is expected to be the highest.
While the thermal pressure of the nebular gas is higher than that 
of the hot gas in the waist region, it is likely that the pressure
of the nebular gas in the lobes is comparable to that of the hot gas.

NGC\,2346 is a larger and more evolved bipolar PN than NGC\,7026.  
No fast wind from the central star of NGC\,2346 has been detected,
as the {\it IUE} observations of the stellar \ion{C}{4} and
\ion{N}{5} lines do not show P\,Cygni profiles \citep{PP91}.
If the central star of NGC\,2346 had a fast stellar
wind in the past, the hot gas resulting from the shocked
fast wind has either cooled or is too rarefied to be detected.

While the presence of a strong fast stellar wind is a prerequisite for the
production of detectable diffuse X-ray emission, the lobe structure of
bipolar PNe may also play an essential role.  NGC\,2346 has an open-lobe
structure and lacks detectable diffuse X-ray emission and NGC\,7026 has a
closed-lobe structure and detectable X-ray emission.  Previous X-ray
observations of bipolar PNe have detected only two other nebulae,
NGC\,7027 and Mz\,3, and both have closed-lobe structures.
It is possible that open lobes
allow shocked-wind to be distributed to a larger volume and the lower
density precludes the detection of X-ray emission.  In the case of 
NGC\,2346, hot gas is not produced currently because of a lack of fast 
stellar wind.  Even if it had a fast wind in the recent past, its open 
lobes were not able to retain the shocked stellar wind to prolong the 
detectability of its diffuse X-ray emission.

X-ray emission is not detected from the central star of 
either NGC\,2346 or NGC\,7026.
The central star of NGC\,2346 is a hot white dwarf with an
A5 companion.  A hot white dwarf's photospheric emission can
reach the soft X-ray energy range, but is easily absorbed by
the nebular gas or foreground interstellar material.  A-type
stars have neither fast stellar winds nor active coronae, and 
therefore are not expected to be X-ray sources.
Even if the white dwarf possesses an accretion disk, the 
temperature of the accretion disk can not be high enough to
produce X-ray emission.
NGC\,7026 has a WC-type central star.
No X-ray emission has ever been detected from a single WC star 
even among the massive Wolf-Rayet stars \citep{Oetal03} because 
of the high opacity in their stellar atmospheres.

\subsection{Hot Gas Content of Bipolar PNe}

To date, only three bipolar PNe have been detected in X-rays.
Among these, Mz\,3 has a symbiotic central star and its X-ray 
emission appears to originate from interactions between its 
bipolar jets and the nebular material \citep{Ketal03}.  
The other two, NGC\,7026 and NGC\,7027, are bona fide 
bipolar PNe.
The X-ray luminosity of NGC\,7026 is one of the highest 
among all PNe detected in X-rays.
Its hot gas mass, $\sim$0.002 $(\frac{d}{1700 pc})^{5/2} 
(\epsilon/0.5)^{1/2} M_\odot$, is also high compared to
other PNe.  
Note however that these values may be overestimated
because the X-ray-emitting lobes may have lower circum-nebular
absorption than the waist region where the extinction was
measured.
The young bipolar PN NGC\,7027 has a lower X-ray luminosity 
and smaller hot gas mass, but a higher hot gas temperature 
and density \citep{Kast01}.
The differences between NGC\,7026 and NGC\,7027 might be 
due to an evolutionary effect, but more positive X-ray 
detections of bipolar PNe of intermediate ages are needed 
to investigate the properties and evolution of their hot gas.

{\it Chandra} has observed five other bipolar PNe but none
were detected.  Based on these non-detections we suggest that
future X-ray observations of bipolar PNe should avoid objects 
that have open bipolar lobes and objects with low velocity 
outflows ($<$300~km~s$^{-1}$).  Instead, observations of 
objects with closed bipolar lobes, to confine hot gas, 
and with kinematic ages of no more than a few thousand years 
would appear best suited for understanding the generation
and evolution of hot gas in bipolar PNe.

\begin{acknowledgements}
This research was supported by the NASA grant NNG04GE63G. 
M.A.G. acknowledges support from the grants AYA~2002-00376 and 
AYA~2005-01495 of the Spanish MEC (cofunded by FEDER funds) and 
the Spanish program Ram\'on y Cajal.

\end{acknowledgements}


\begin{thebibliography}{}

\bibitem[Acker \& Neiner(2003)]{AN03} 
         Acker, A., \& Neiner, C.\ 2003, \aap, 403, 659 

\bibitem[Aller(1976)]{Aller76} 
         Aller, L.~H.\ 1976, Mem.\ Soc.\ R.\ Sci.\ Li\`ege, 9, 271 
 
\bibitem[Anders \& Grevesse(1989)]{AG89} 
         Anders, E., \& Grevesse, N.\ 1989, Geochimica et Cosmochimica 
         Acta 53, 197

\bibitem[Arias et al.(2001)]{Aetal01} 
         Arias, L., Rosado, M., Salas, L., \& Cruz-Gonz\'alez, I.\ 2001, 
         \aj, 122, 3293 

\bibitem[Balucinska-Church \& McCammon(1992)]{BM92} 
         Balucinska-Church, M., \& McCammon, D.\ 1992, ApJ, 400, 699

\bibitem[Bohlin et al.(1978)Bohlin, Savage \& Drake]{BSD78}
         Bohlin, R.\ C., Savage, B.\ D., \& Drake, J.\ F.\ 1978, \apj, 
         224, 132

\bibitem[Chu et al.(2001)]{CGGWK01} 
         Chu, Y.-H., Guerrero, M.\ A., Gruendl, R.\ A., Williams, R.\ M., 
         \& Kaler, J.\ B.\ 2001, \apjl, 553, L69 

\bibitem[Ciardullo et al.(1999)]{Cetal99} 
         Ciardullo, R., Bond, H.\ E., Sipior, M.\ S., Fullton, L.\ K., 
         Zhang, C.-Y., \& Schaefer, K.\ G.\ 1999, \aj, 118, 488 

\bibitem[Cuesta, Phillips, \& Mampaso(1996)]{CPM96}
         Cuesta, L., Phillips, J.\ P., \& Mampaso, A.\ 1996, \aap, 313, 243

\bibitem[Frank \& Mellema(1994)]{FM94} 
         Frank, A., \& Mellema, G.\ 1994, \apj, 430, 800 

\bibitem[Guerrero et al.(2005a)]{GCGM05} 
         Guerrero, M.\ A., Chu, Y.-H., Gruendl, R.\ A., \& Meixner, M.\ 2005a, 
         \aap, 430, L69 

\bibitem[Guerrero et al.(2005b)]{GCG05} 
         Guerrero, M.\ A., Chu, Y.-H., Gruendl, R.\ A.\ 2005b,
         in Planetary Nebulae as Astronomical Tools, Eds.\
	 R.\ Szczerba, G.\ Stasinska, and S.\ K.\ Gorny, 
	 AIP Conf.\ Proc.\ 804, 157

\bibitem[Hamann et al.(2005)Hamann, Todt \& Graefener]{Hetal05} 
         Hamann, W.-R., Todt, H., \& Graefener, G.\ 2005, in Planetary Nebulae
         as Astronomical Tools, Eds.\ R.\ Szczerba, G.\ Stasinska, and 
         S.\ K.\ Gorny, AIP Conf.\ Proc.\ 804, 153 
 
\bibitem[Hyung \& Feibelman(2004)]{HF04} 
         Hyung, S., \& Feibelman, W.\ A.\ 2004, \apj, 614, 745 

\bibitem[Kaastra \& Mewe(1993)]{KM93}
        Kaastra, J.\ S., \& Mewe, R.\ 1993, Legacy, 3, 16, HEASARC, NASA

\bibitem[Kastner et al.(2003)]{Ketal03} 
         Kastner, J.\ H., Balick, B., Blackman, E.\ G., Frank, A., Soker, 
         N., Vrt{\'{\i}}lek, S.\ D., \& Li, J.\ 2003, \apjl, 591, L37 

\bibitem[Kastner et al.(2001)]{Kast01} 
         Kastner, J.\ H., Vrtilek, S.\ D., \& Soker, N.\ 2001, \apjl, 550, 
         L189 

\bibitem[Koesterke(2001)]{K01}
         Koesterke, L.\ 2001, Ap\&SS, 275, 41

\bibitem[Kwitter \& Henry(2001)]{KH01} 
         Kwitter, K.\ B., \& Henry, R.\ B.\ C.\ 2001, \apj, 562, 804 

\bibitem[Kwok(1983)]{K83} 
         Kwok, S.\ 1983, in IAU Symposium 103, Planetary Nebulae, ed. D.\ 
         R.\ Flower (Dordrecht: Reidel), 293

\bibitem[Liedahl et al.(1995)Liedahl, Osterheld, \& Goldstein]{LOG95}
        Liedahl, D.\ A., Osterheld, A.\ L., \& Goldstein, W.\ H.\ 1995,
        \apj, 438, L115

\bibitem[L\'opez(2003)]{L03}
         L\'opez, J.\ A.\ 2003, Rev.\ Mex.\ Astron.\ y Astrof., 13, 139

\bibitem[M\'endez(1978)]{M78} 
         M\'endez, R.\ H.\ 1978, \mnras, 185, 647 

\bibitem[M\'endez \& Niemela(1981)]{MN81} 
         M\'endez, R.\ H., \& Niemela, V.\ S.\ 1981, \apj, 250, 240 

\bibitem[Oskinova et al.(2003)]{Oetal03}
         Oskinova, L.~M., Ignace, R., Hamann, W.-R., Pollock, A.~M.~T., 
         \& Brown, J.~C.\ 2003 \aap, 402, 755

\bibitem[Patriarchi \& Perinotto(1991)]{PP91} 
         Patriarchi, P., \& Perinotto, M.\ 1991, \aaps, 91, 325 

\bibitem[Solf \& Weinberger(1984)]{SW84} 
         Solf, J., \& Weinberger, R.\ 1984, \aap, 130, 269 

\bibitem[Su et al.(2004)]{Setal04} 
         Su, K.\ Y.\ L., et al.\ 2004, \apjs, 154, 302 

\bibitem[Terzian(1993)]{T97} 
         Terzian, Y.\ 1997, IAU Symp.~180: Planetary Nebulae, 180, 29 
 
\bibitem[Weaver et al.(1977)]{Wetal77}
         Weaver, R., McCray, R., Castor, J., Shapiro, P., \& Moore, R.\ 1977,
         \apj, 218, 377

\bibitem[Walsh et al.(1991)]{WMW91} 
         Walsh, J.\ R., Meaburn, J., \& Whitehead, M.\ J.\ 1991, 
         \aap, 248, 613 

\bibitem[Zhekov \& Perinotto(1996)]{ZP96}
         Zhekov, S.\ A., \& Perinotto, M.\ 1996, \aap, 309, 648

\bibitem[Zuckerman \& Gatley(1988)]{ZG88} 
         Zuckerman, B., \& Gatley, I.\ 1988, \apj, 324, 501 


\end{thebibliography}
\end{document}